\documentclass[aps,pra,twocolumn,superscriptaddress]{revtex4-2}

\usepackage{graphicx}% Include figure files
\usepackage{amsmath}
\usepackage{amssymb}
\usepackage{empheq}
\usepackage{physics}
\usepackage{xcolor}

\begin{document}

%\preprint{}

\title{Fast and accurate modelling of Kerr-Brillouin combs in Fabry-Perot resonators}


%\author{}
%\thanks{}
%\affiliation{}


\author{Matteo Conforti}\email{matteo.conforti@univ-lille.fr}
\author{Thomas Bunel}
\affiliation{University of Lille, CNRS, UMR 8523-PhLAM Physique des Lasers, Atomes et Molécules, F-59000, Lille, France}
\author{Auro M. Perego}
\affiliation{Aston Institute of Photonic Technologies, Aston University, Birmingham, B4 7ET, UK}
\author{Arnaud Mussot}
\affiliation{University of Lille, CNRS, UMR 8523-PhLAM Physique des Lasers, Atomes et Molécules, F-59000, Lille, France}

%\author{Matteo Conforti}\email{matteo.conforti@univ-lille.fr}
%\affiliation{University of Lille, CNRS, UMR 8523-PhLAM Physique des Lasers, Atomes et Molécules, F-59000, Lille, France}

\date{\today}

\begin{abstract}
%We propose a new mean-field equation for the modelling of Fabry-Perot resonators filled by a dispersive medium featuring Brillouin and Kerr nonlinearity. It is derived  starting from a unified approach to cascaded Brillouin scattering and four-wave mixing, which consists in two coupled nonlinear Schr\"odinger  equations for the forward and backward propagating fields and a single equation describing the acoustic oscillation. Under the usual assumptions used to derive  mean-field models (good finesse, weak non-linearity and dispersion), we show that our equation is in excellent agreement with the original system.  Our model permits to develop physical insights thanks to its simple and elegant mathematical structure. As a relevant example, we derive a simple expression for the growth rate of harmonic perturbations of the steady states. In addition, our equation enables fast and accurate numerical simulations by means of standard Fourier split step methods. We show the power of our method by simulating the generation of frequency combs in state-of-the-art high-Q fiber Fabry-Perot resonators. 
We introduce a new mean-field equation for modeling Fabry-Perot resonators filled with a dispersive medium exhibiting both Brillouin and Kerr nonlinearities, e.g. an optical fiber. This model is derived from a unified framework that accounts for Brillouin scattering and four-wave mixing. It involves two coupled nonlinear Schrödinger equations for the forward and backward propagating fields, alongside a single equation governing the acoustic oscillation. Under the standard assumptions for the mean-field approach (high finesse, weak nonlinearity, and weak dispersion) we demonstrate that our model closely matches the original system. The simplified and elegant mathematical structure of our equation provides valuable physical insights. As a key example, we derive an expression for the growth rate of harmonic perturbations to the steady states. Additionally, our model facilitates fast and accurate numerical simulations using standard Fourier split-step methods. We highlight the effectiveness of this approach by simulating frequency comb generation in state-of-the-art high-Q fiber Fabry-Perot resonators.
\end{abstract}

%\keywords{}

\maketitle

\section{Introduction}

Optical frequency combs (OFCs), periodic trains of phase-locked laser pulses with spectra consisting of equally spaced lines, have transformed precision metrology and significantly influenced various applications that require precise measurement of absolute frequency \cite{Diddams2020}. First delivered by bulky mode-locked lasers, OFCs has been successfully generated by pumping passive high-finesse microresonators featuring Kerr nonlinearity \cite{Pasquazi2018}. In these devices, the recirculation of distinct, well-defined pulses, such as temporal cavity solitons \cite{Leo2010,Herr2013, Bunel2024} or switching waves \cite{Bunel2024SW,Anderson2022,Macnaughtan2023,Xue2015,Xue2016}, or multi-peaked temporal structures like Turing patterns (primary combs) \cite{Pfeifle2015} or soliton crystals \cite{Cole2017}, enables the generation of broadband, stable frequency combs.
In particular, OFC delivered by microresonators, or microcombs, showed impressive performances in optical communications, spectroscopy and many others \cite{Sun2023}. Even if the majority of the literature is devoted to travelling-wave resonators (microring, microsphere...), quite recently the attractiveness of standing-wave (Fabry-Perot) resonators has been unveiled \cite{Obrzud2017,Wildi2023}. In particular, fiber FP  (FFP) resonators combine the advantages of microresonator and fiber ring cavities. They namely exhibit a high-quality factor, compact design, free-spectral range (FSR) in the gigahertz range, simplified light coupling, and ease of implementation within fiber systems \cite{Musgrave2023}. From the very first studies of FFP resonators, the impact of Stimulated Brillouin Scattering (SBS) \cite{Boyd} has been identified \cite{Braje2009}. Indeed, SBS is the first nonlinear effect to show up when pumping with continuous wave (CW) or quasi CW pulses (provided that SBS gain overlaps a cavity resonance). Beside being useful for stabilizing  Kerr microcombs \cite{Nie2024,Jia2020}, SBS can lead to the generation of Kerr-Brillouin combs in both ring and FP configuration \cite{Bai2021,Zhang2024,Zhang2023,Buttner2014,Lucas2022,postdeadline,BunelNatPhot}.

Different methods are used to model the nonlinear dynamics of OFCs. Coupled mode theory (CMT) is traditionally used by the microresonator community. Several first-order ordinary differential equations coupled by four-wave mixing terms, describe the temporal evolution of the modes' amplitudes \cite{Matsko2005,Savchenkov2016,Chembo2010,Chembo2013,Hansson2014}. The Ikeda map, an iterative mapping developed in the late 1970s in the context of optical bistability in passive resonators, has since become a foundational model for the fiber optics community \cite{Ikeda1979,Blow1984,McLaughlin1985}. In the good cavity limit (aka unfiorm- or mean-field limit), an elegant equation named after L. Lugiato and R. Lefever (LLE) has been derived in the context of spatial pattern formation \cite{Lugiato1987}, consisting in a driven and damped nonlinear Schr\"odinger equation \cite{Nozaki1985,Barashenkov1996}. LLE and its generalizations \cite{Coen2013,Conforti2017,Kartashov2017} have become the workhorse of the cavity nonlinear dynamics \cite{Castelli2017}, due to its simple mathematical structure and its simple and efficient numerical solution. Moreover, LLE has been derived starting from different general models: Maxwell-Bloch  equations \cite{lugiato_prati_brambilla_2015}, Ikeda map \cite{Haelterman1992}, CMT \cite{Chembo2013}, making it an unifying tool for different resonator platforms.

Specific to Fabry-Perot resonators, the presence of two counter-propagating field makes the modelling much more challenging. For instance, it is not possible to derive a simple analogous of the Ikeda map, because of the multiple time delays arising from the exchange of information between the fields propagating in opposite directions \cite{Ikeda1985}. Modelling of FP resonators thus entails the use of coupled propagation equations and boundary conditions at the mirrors \cite{lugiato_prati_brambilla_2015,Lugiato1988,Firth1981,Firth2021,Ziani2024}. Quite recently, CMT \cite{Obrzud2017} and LLE \cite{Cole2018} have been generalized to FP geometry (FPLLE).  The modeling of SBS encounters similar difficulties, because the radiation scattered by the acoustic wave is phase-matched in the backward direction. A few years ago, a model for the description of cascaded SBS and four-wave mixing has been proposed \cite{Dong2016}. It consists in three coupled wave equations (CWE): two for the forward and backward optical fields and one for the acoustic wave. The interest of this approach to frequency combs is that all the Stokes  and anti-Stokes lines emerging from SBS and FWM arise naturally. For example, to model a Brillouin comb in a FP resonator up to only third-order Stoke waves, 14 coupled equations are necessary \cite{Ogusu2002}. Despite being a very general model, the analysis and numerical solution of CWE remains an hard task \cite{Sun2019}. In addition, the complexity of this model obscures any underlying physical insights. Alternative approaches based on CMT have been proposed for the description of Kerr-Brillouin combs in both ring \cite{Bai2021,Zhang2024,Zhang2023}  and FP resonators \cite{Nie2024}, but due to strong assumptions only a limited number of interaction can be accurately described by these models.

In this paper we derive a single mean-field equation which permits to accurately model the generation of broadband Kerr-Brillouin combs generated by cascaded SBS and multiple FWM. Its form is similar to the FPLLE generalized  with the inclusion of a nonlocal term accounting for the Brillouin response. We show that the simple and elegant structure permits to obtain strong physical insights which may be hidden in the original equations. 
Moreover, its numerical solution is extremely efficient and permits to drastically reduce the computational time up to four orders of magnitude. The model introduced here extends our preliminary formulation \cite{BunelNatPhot} to include Brillouin cascade to all orders. Despite being more general, our novel equation  features a simpler and more straightforward mathematical form.
We show the power of our approach by simulating the generation of wideband Kerr-Brillouin OFCs in FFP cavities, as observed in recent experiments \cite{postdeadline,BunelNatPhot}.

\section{Coupled wave description}

\begin{figure}
\includegraphics[width=0.9\columnwidth]{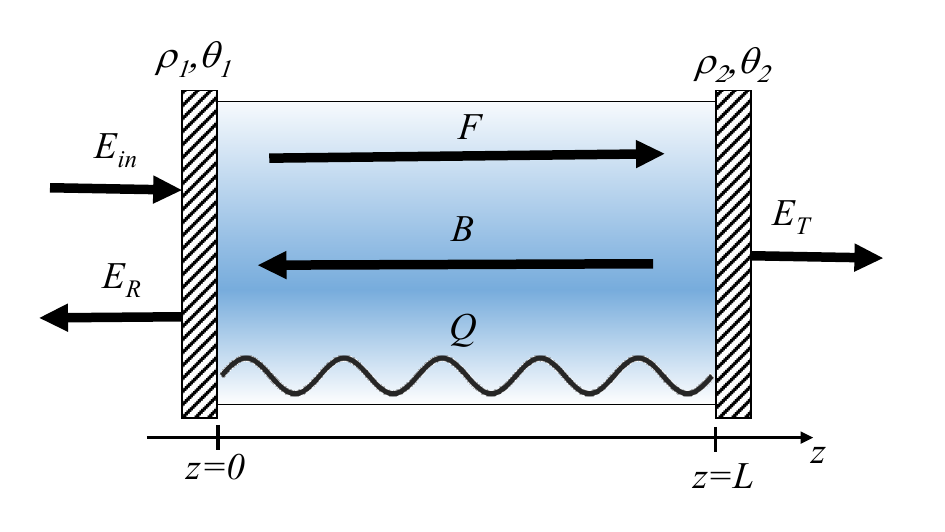}
\caption{ Scheme of the FP resonator considered in the text. $F,B$: Forward and backward electric field envelopes; $Q$: acoustic wave; $E_{in}$: input field; $E_R,E_T$: reflected and transmitted fields; $\rho_i,\theta_i$: mirrors' reflectivities and transmissivities.\label{scheme}}
\end{figure}

We consider a Fabry-Perot cavity of length $L$ filled by a dispersive  medium featuring Kerr and Brillouin nonlinearities (e.g. a single-mode optical fiber), as sketched in Fig. \ref{scheme}. A pump field $E_{in}$ enters at $z = 0$ through a mirror of reflectivity $\rho_1$ and drives forward $F(z, t)$ and backward $B(z, t)$ fields in the cavity.
A transmitted field $E_{T}$ exits the cavity through the second mirror of reflectivity $\rho_2$ at $z = L$. The evolution equations for the slowly varying forward $F(z,t)$ and backward $B(z,t)$ optical fields and the acoustic wave envelopes $Q(z,t)$ read \cite{Dong2016}:
\begin{align}
\nonumber \frac{\partial F}{\partial z}+\beta_1\frac{\partial F}{\partial t} 
+ i\frac{\beta_2}{2}\frac{\partial^2 F}{\partial t^2}&= i\gamma\left(|F|^2+X|B|^2\right)F\\
\label{propF}&+\frac{g_B}{2A_{eff}}QB,\\
\nonumber-\frac{\partial B}{\partial z}+\beta_1\frac{\partial B}{\partial t} 
+ i\frac{\beta_2}{2}\frac{\partial^2 B}{\partial t^2}&=i\gamma\left(|B|^2+X|F|^2\right)B\\
\label{propB}&-\frac{g_B}{2A_{eff}}Q^*F,\\
\left(\frac{\partial^2}{\partial t^2}+\Gamma_B\frac{\partial}{\partial t}+\Omega_B^2\right)Q&=i\Omega_B\Gamma_B FB^*,\label{Q}
\end{align}
where $v_g=\beta_1^{-1}$, $\beta_2$ and $\gamma$ (m$^{-1}$W$^{-1}$) are the group velocity, the group velocity dispersion (GVD) and the nonlinear Kerr coefficient. The parameter $X$ describes the cross-phase modulation (XPM) and is taken to be $X=2$ \cite{agrawal2006,Firth2021}. Here $g_B$ (m/W) is the Brillouin gain, $A_{eff}$ is the effective area of the fiber, $\tau_B=1/\Gamma_B$ and $\Omega_B$ are the Brillouin lifetime and frequency shift. The difference of a factor 4 in the right hand side of Eq. (\ref{Q}) with respect to the original formulation \cite{Dong2016} depends on the definition of the complex envelopes \cite{envelopes}. The convention employed here is consistent with the fiber-optic community standard \cite{agrawal2006}, which permits to employ the usual values of the nonlinear coefficients. We have also included GVD, which is essential for describing Kerr OFCs, whether generated by  modulation instability (MI), solitons (bright or dark), or switching waves.
The optical fields are normalized in such a way that their modulus square gives the instantaneous power. 
Equation (\ref{Q}) neglects propagation of phonons (no spatial derivatives), which are assumed to be heavily dampened.  Both forward and backward propagating waves are included due to the second-order time derivative in the acoustic wave equation \cite{Dong2016}.
%Equation (\ref{Q}) assumes that phonons are heavily damped and propagate only over short distances, which justifies neglecting spatial derivatives. However, the second derivative with respect to time is retained. Consequently, the acoustic wave equation describes a driven and damped harmonic oscillator with natural frequencies $\pm \Omega_B$. The positive and negative frequencies correspond to forward and backward acoustic waves, respectively \cite{Dong2016}.%
%

Coupled wave equations (\ref{propF},\ref{Q}) are supplemented with the following boundary conditions at input ($z=0$) and output ($z=L$) mirrors:
\begin{align}
F(0,t)&=\theta_1 E_{in} + \rho_1 B(0,t),\label{BC1}\\
B(L,t)&= \rho_2e^{i\phi_0} F(L,t),\label{BC2}
\end{align}
where $\theta_{1,2}$ ($\rho_{1,2}$) are the mirrors transmissivity (reflectivity),  and  the phase $\phi_0$ accounts for the phase acquired during the propagation over one roundtrip ($2\beta_0 L$) and any possible contribution from the mirrors. All the losses are lumped in the coefficients $\rho_{1,2}$. We define $\delta=2\pi m_0-\phi_0$ as the phase detuning between the driving field and the nearest resonance (indexed by the integer $m_0$).
Before focussing on the formulation of the mean-field equation, we derive some useful results from Eqs. (\ref{propF}-\ref{BC2}).

\subsection{Steady states}
Equations (\ref{propF}-\ref{Q}) admit the following CW (time-independent) solution:
\begin{align}
        \label{stazF}F(z) &= F_0 e^{i\gamma (1+X_{eff}\rho_2^2)P z} , \\
        \label{stazB}B(z) &= B_0 e^{-i\gamma (\rho_2^2+X_{eff})P z} ,\\
        Q(z)&=\frac{i\Gamma_B}{\Omega_B}F(z)B(z)^*
\end{align}
where $P=|F_0|^2$ and we have defined an effective XPM coefficient  $X_{eff}=X+\gamma_B/\gamma$, with 
$\gamma_B=\frac{g_B \Gamma_B}{2A_{eff}\Omega_B}$. Brillouin nonlinearity for CW fields at the same frequency does not lead to energy exchange between forward and backward waves, but only to a phase-dependent interaction which enhances the effect of Kerr nonlinearity.
Using Eqs. (\ref{BC1},\ref{BC2}) we find the complex amplitudes of the steady states:\textit{
    \begin{eqnarray}
        F_0 &=& \frac{\theta_1 E_{in}}{1-\rho_1\rho_2 \exp{[i(\phi_0+\phi_{NL})}]}\;,\\
        B_0 &=& \rho_2 \exp{[i(\phi_0+\phi_{NL})]}F_0.
    \end{eqnarray}}
where the nonlinear phase is given by~:
\begin{equation}
    \phi_{NL} = \gamma (1+\rho_2^2)(1+X_{eff})L P.
\end{equation}
The input power $P_{in}=|E_{in}|^2$ can be written as a function of the intracavity forward power $P$ as follows:
\begin{equation}\label{eq:SS_power}
    P_{in} = \frac{P}{\theta_1^2} \left( 1 + (\rho_1\rho_2)^2 -2\rho_1\rho_2 \cos(\phi_0 + \phi_{NL})\right).
\end{equation}
It is worth noting that the effect of SBS on the steady response  is to increase the cross-phase modulation coefficient, without changing the shape of the response which is identical to the pure Kerr case \cite{Ziani2024}.

\subsection{Brillouin gain spectrum}
In order to derive the Brillouin gain spectrum from Eqs. (\ref{propF}-\ref{Q}),  we  neglect Kerr  and dispersion effects, and consider a  forward pump at frequency $\omega_p$  and a backward signal at $\omega_s$. For pump durations much greater than the Brillouin lifetime $\tau_B=1/\Gamma_B$, the acoustic wave reaches quickly its steady steate, allowing to neglect the temporal dependence and simplifying considerably the dynamics \cite{agrawal2006}. We perform the following change of variables:
\begin{align}
F(z,t)&=\tilde F(z,t)e^{-i\omega_p (t-\beta_1 z)},\\
B(z,t)&=\tilde B(z,t)e^{-i\omega_s (t+\beta_1 z)},\\
Q(z,t)&=\tilde Q(z)e^{-i(\omega_p-\omega_s )t}=\tilde Q(z)e^{-i\omega t}.
\end{align}

By defining the powers $P_F=|F|^2$ and $P_B=|B|^2$ we get the usual coupled-wave description of SBS in propagation \cite{agrawal2006}:
\begin{align}
\label{PF}\frac{\partial P_F}{\partial z}+\beta_1\frac{\partial P_F}{\partial t} 
&=-\frac{G_B(\omega)}{A_{eff}}P_F P_B,\\
\label{PB}-\frac{\partial P_B}{\partial z}+\beta_1\frac{\partial P_B}{\partial t} 
&=\frac{G_B(\omega)}{A_{eff}}P_F P_B,
\end{align}
 where $\omega=\omega_p-\omega_s$. The Brillouin gain has a double Lorenzian shape:
\begin{equation}
 G_B(\omega)=g_B\frac{\omega\Omega_B\Gamma_B^2}{(\Omega_B^2-\omega^2)^2+\omega^2\Gamma_B^2}. \label{gain_brillouin_2}
 \end{equation}
For $\omega\approx\Omega_B$ we recover the conventional single Lorenzian shape:
\begin{equation}
 G_B(\omega)\approx g_B\frac{(\Gamma_B/2)^2}{(\omega-\Omega_B)^2+(\Gamma_B/2)^2}.
\end{equation}

\section{Mean-field model: derivation}
In order to derive a mean-field model from Eqs. (\ref{propF}-\ref{Q}), we follow the original approach by Lugiato \emph{et al.} developed in the context of laser cavities \cite{lugiato_prati_brambilla_2015}, which we have recently specialized to Kerr fiber FP resonators \cite{Ziani2024}. The main steps are: (i) changing variables to enforce periodic boundary conditions; (ii) applying the good-cavity (also known as mean-field or uniform field) approximation; and (iii) deriving a partial differential equation from the modal equations. The key novelty of our approach is incorporating the acoustic wave into a single mean-field equation through a carefully defined Brillouin response.
\subsection{Modal equations}
As in Refs. \cite{Lugiato1988,lugiato_prati_brambilla_2015,Ziani2024}, we use the following change of variables:
\begin{align}
\label{T1}f(z,t)&=e^{\frac{z-L}{L}\left(\ln \rho_1-i\frac{\delta}{2}\right)-\nu z}F(z,t)+\frac{\theta_1}{\rho_1}e^{i\frac{\delta}{2}}\frac{z-L}{2L}E_{in},\\
\label{T2}b(z,t)&=e^{-\frac{z}{L}\left(\ln \rho_2-i\frac{\delta}{2}\right)-\nu z}e^{i\frac{\delta}{2}}B(z,t)-\frac{\theta_1}{\rho_1}e^{i\frac{\delta}{2}}\frac{z-L}{2L}E_{in},
\end{align}
with $\nu=\frac{1}{2 L}\ln\frac{\rho_1}{\rho_2}$ accounting for possible differences in the mirrors. We consider a CW pump $E_{in}=const.$, but the derivation is easily extended to pulsed synchronous pumping, as showed in \cite{Ziani2024}.
The boundary conditions Eqs. (\ref{BC1},\ref{BC2}) for the new variables are simplified to:
\begin{align}
f(0,t)&=b(0,t),\label{BC1new}\\
b(L,t)&=f(L,t).\label{BC2new}
\end{align}
%
%However, the simplification of the boundary condition  on the one hand, implies an increase of the complexity of the propagation equations for $f,b$ on the other end. 
However, while simplifying the boundary condition reduces complexity in one aspect, it increases the complexity of the propagation equations for $f$ and $b$.
From now on we thus restrict the analysis to good cavities by assuming $\rho_{1,2}\rightarrow 1$, $\theta_{1,2}\rightarrow 0$, $\delta\rightarrow 0$, $\nu\rightarrow 0$. At zero-th order we have $F\approx f$ and $B\approx b$. From Eqs.~(\ref{T1}-\ref{T2}) we calculate $\partial_z F, \partial_t F, \partial_z B ,\partial_t B$ as a function of $f,b$ and their derivatives.
We truncate the obtained expressions at first order in $\rho_{1,2}$ and $\delta$, then substitute them into Eqs.~(\ref{propF},\ref{propB}). Since dispersion and nonlinearity are treated as first-order corrections—consistent with the assumptions underlying the propagation equations (\ref{propF},\ref{propB})—we apply zero-order expansions to the dispersive and nonlinear terms. These approximations significantly simplify the propagation equations as follows:
%\begin{widetext}
\begin{align}
\nonumber \frac{\partial f}{\partial z}&-\frac{\theta_1}{2L}E_{in}-\left(\frac{\ln\rho_1\rho_2}{2L}-i\frac{\delta}{2L}\right)f+\beta_1\frac{\partial f}{\partial t} 
+ i\frac{\beta_2}{2}\frac{\partial^2 f}{\partial t^2} \\
&=i\gamma \left(|f|^2+X|b|^2\right)f+\frac{g_B}{2A_{eff}}Qb, \label{propfMF}\\
\nonumber -\frac{\partial b}{\partial z}&-\frac{\theta_1}{2L}E_{in}-\left(\frac{\ln\rho_1\rho_2}{2L}-i\frac{\delta}{2L}\right)b+\beta_1\frac{\partial b}{\partial t} 
+ i\frac{\beta_2}{2}\frac{\partial^2 b}{\partial t^2}\\
&=i\gamma \left(|b|^2+X|f|^2\right)b-\frac{g_B}{2A_{eff}}Q^*f \label{propbMF}.
\end{align}
%\end{widetext}
%
%The next step to unite Eqs. (\ref{propfMF},\ref{propbMF}) into a single equation involves expressing the fields as a sum of the ideal cavity's modes. 
Next, to transform Eqs.~(\ref{propfMF},\ref{propbMF}) into a single evolution equation, we express the forward and backward fields as superpositions of the eigenmodes of the ideal Fabry–Perot cavity. These modes form a complete basis that naturally incorporates the boundary conditions, allowing us to capture the spatial structure of the field while reducing the problem to a set of coupled-mode equations. The modes of the loss-less, empty and undriven cavity [$\beta_2=\delta=E_{in}=\gamma=g_B=0$, $\rho_1=\rho_2=1$ in Eqs. (\ref{propfMF},\ref{propbMF})] can be written as:
\begin{align}
f_m(z,t)=e^{-i\omega_m t }e^{i k_m z }=e^{-i\omega_m (t - \beta_1 z)},\\
b_m(z,t)=e^{-i\omega_m t }e^{-i k_m z }=e^{-i\omega_m(t + i \beta_1 z)},
\end{align}
where
\begin{equation}
\label{modes} \omega_m=\frac{m\pi}{\beta_1L},\;\;k_m=\beta_1\omega_m=\frac{m\pi}{L}.
\end{equation}
The fields in Eqs. (\ref{propfMF},\ref{propbMF}) are expanded as a sum of cavity modes, allowing for a slow variation in the modal amplitudes :
\begin{align}
\label{expanF}f(z,t)=\sum_m a_m(t)e^{-i\omega_m (t - \beta_1 z)},\\
\label{expanb}b(z,t)=\sum_m a_m(t)e^{-i\omega_m(t + i \beta_1 z)}.
\end{align}
It is worth noting that in spite of having two propagation equations, we obtain a single set of modal amplitudes, a peculiarity of standing-wave resonators. 

The expansions Eqs. (\ref{expanF},\ref{expanb}) are substituted in Eq. (\ref{propfMF}) and the resulting equations are projected over the forward modes by multiplying by $\frac{1}{2L}\exp[i\omega_n (t - \beta_1 z)]$ and integrating in $z$ in the interval $(-L,L)$. For each mode indexed by $n$, the following equation is obtained:
\begin{align}
\nonumber \beta_1 \dot a_n -\left(\frac{\ln\rho_1\rho_2}{2L} - i\frac{\delta}{2L}\right)a_n -\frac{\theta_1}{2L}E_{in}\delta_{n0} \\+i\frac{\beta_2}{2}\left(\ddot a_n -2i\omega_n\dot a_n -\omega_n^2a_n\right)=i\gamma \mathcal{K}_n^{f}+ \frac{g_B}{2A_{eff}} \mathcal{B}_n^{f}.  \label{modal1}
\end{align}
Inserting the expansions Eqs. (\ref{expanF},\ref{expanb}) in Eq. (\ref{propbMF}), and projecting over the backward modes (multiplication by $\frac{1}{2L}\exp[i\omega_n (t + \beta_1 z)]$ and integratation in $z$) gives a similar result:
\begin{align}
\nonumber \beta_1 \dot a_n -\left(\frac{\ln\rho_1\rho_2}{2L} - i\frac{\delta}{2L}\right)a_n -\frac{\theta_1}{2L}E_{in}\delta_{n0} \\+i\frac{\beta_2}{2}\left(\ddot a_n -2i\omega_n\dot a_n -\omega_n^2a_n\right)=i\gamma \mathcal{K}_n^{b}- \frac{g_B}{2A_{eff}} \mathcal{B}_n^{b}.\label{modal1b}
\end{align}
Here, $\delta_{n0}$ is the Kroneker delta and the nonlinear forward/backward Kerr and Brillouin contributions $\mathcal{K}_n^{f/b}$ and $\mathcal{B}_n^{f/b}$ are described below.

\emph{Kerr term} -- The Kerr term in the equation for the forward field in Eq. (\ref{propfMF}) is written as a combination of modes  Eqs. (\ref{expanF},\ref{expanb}) and projected over the $n$-th forward mode, to obtain \cite{Cole2017,Ziani2024}
%\begin{equation}\label{K}
%(|f|^2+X|b|^2)f=\sum_{n,n',n''}a_{n'}a_{n''}^*a_{n-n'+n''}e^{-i\omega_n(t-\beta_1z)}+X\sum_{n,n',n''}a_{n'}a_{n''}^*a_{n+n'-n''}e^{i\omega_n \beta_1 z}e^{-i(\omega_n +2(\omega_{n'}-\omega_{n''}))t}
%\end{equation}
%By projecting Eq. (\ref{K}) over the forward modes  %i.e. by applying the following operator $$\frac{1}{2L}\int_{-L}^L(.)e^{i\omega_p(t-\beta_1z)}dz,$$
%we get
\begin{align}\label{KFn}
\nonumber \mathcal{K}^f_{n}=&\frac{1}{2L}\int_{-L}^L(|f|^2+X|b|^2)fe^{i\omega_n(t-\beta_1z)}dz=\\
\nonumber &\sum_{n',n''}a_{n'}a_{n''}^*a_{n-n'+n''}\\
&+X\sum_{n',n''}a_{n'}a_{n''}^*a_{n+n'-n''}e^{-i2(\omega_{n'}-\omega_{n''})t}.
\end{align}
The same result is obtained for the Kerr term in the backward equation (\ref{propbMF}):
\begin{align}
\nonumber \mathcal{K}^b_{n}=&\frac{1}{2L}\int_{-L}^L(|b|^2+X|f|^2)be^{i\omega_n(t+\beta_1z)}dz=\mathcal{K}^f_{n}.\; \blacksquare
\end{align}

\emph{Brillouin term} -- We separate the variables in the acoustic wave, expanding the spatial dependence over the cavity modes while keeping the temporal dependence in a general form  :
\begin{equation}
\label{expanQ}Q(z,t)=\sum_{m}Q_m(t)e^{i k_m z}.
\end{equation}
We substitute Eq. (\ref{expanQ}) into Eq. (\ref{Q}), expand the nonlinear term over the cavity modes, and then gather the terms oscillating with the same wavenumber $k_m$, to obtain:
\begin{align}
\nonumber \left(\frac{\partial^2}{\partial t^2}+\Gamma_B\frac{\partial}{\partial t}+\Omega_B^2\right)Q_m(t)=
\\i\Gamma_B\Omega_B\sum_n a_na_{m-n}^*e^{-i(2\omega_n-\omega_m)t}.\label{Qm}
\end{align}
For each value of $m$, Eq. (\ref{Qm}) describes a forced harmonic oscillator, thus we seek a solution of the following form:
\begin{equation}
\label{expanQm}Q_m(t)=\sum_{p}q_{mp}(t)e^{-i \Omega_pt}.
\end{equation}
We substitute Eq. (\ref{expanQm}) in Eq. (\ref{Qm})  and assume that $q_{mp}$ varies slowly over time with respect to $\Omega_p$, yielding:
\begin{align}
\nonumber\sum_p\left(\Omega_B^2-\Omega_p^2-i\Omega_p\Gamma_B\right)q_{mp}(t)e^{-i\Omega_p t}=\\
i\Omega_B\Gamma_B\sum_{n}a_na_{m-n}^*e^{-i(2\omega_n-\omega_m)t}.
\end{align}
By collecting terms oscillating at the same frequency, we find $\Omega_p=2\omega_n-\omega_m$ and
\begin{equation}
q_{m,2n-m}=iH_B(2\omega_n-\omega_m)a_na_{m-n}^*,
\end{equation}
where we have defined the Brillouin response in the frequency domain as:
\begin{equation}\label{Hb}
H_B(\omega_n)=H_B(k_n/\beta_1)=\frac{\Omega_B\Gamma_B}{\Omega_B^2-\omega_n^2-i\omega_n\Gamma_B}.
\end{equation}
The acoustic wave envelope has thus the following expression:  
\begin{equation}\label{Qnm}
Q(z,t)=i\sum_{n,m}a_na_{m-n}^*H_B(2\omega_n-\omega_m)e^{-i(2\omega_n-\omega_m)t}e^{i k_m z}.
\end{equation}
The forward Brillouin term in Eq. (\ref{propfMF}) is first written as a combination of modes thanks to Eqs. (\ref{expanF},\ref{expanb},\ref{Qnm})
%\begin{align}
%\nonumber Qb=i\sum_{n,m,p}H_B(2\omega_n-\omega_m)a_nb_{m-n}^*b_pe^{-i(2\omega_n-\omega_m+\omega_p)t}e^{i (k_m-n_p) z}.
%\end{align}
%
and then projected over the forward modes to obtain:
\begin{align}
&\nonumber\mathcal{B}^f_{n}=\frac{1}{2L}\int_{-L}^LQbe^{i\omega_n(t-\beta_1z)}dz=\\
&i\sum_{n',n''}a_{n'}a_{n''}a^*_{n-n'+n''}H_B(2\omega_{n'}-\omega_{n''}-\omega_n)e^{-i2(\omega_{n'}-\omega_n)t}.\label{NBf}
\end{align}
For the backward Brillouin term, by following a similar procedure starting from Eq. (\ref{propbMF}),  we obtain:% is proportional to
%\begin{align}
%\nonumber Q^*f=-i\sum_{n,m,p}a_n^*b_{m-n}a_pH_B^*(2\omega_n-\omega_m)e^{i(2\omega_n-\omega_m-\omega_p)t}e^{-i (k_m-n_p) z}.
%\end{align}
%%
%By projecting over the backward modes we get
\begin{align}
\nonumber &\mathcal{B}_n^{b}=\frac{1}{2L}\int_{-L}^LQ^*fe^{i\omega_n(t+\beta_1z)}dz=\\
&-i\sum_{n',n''}a_{n'}^*a_{n''}a_{ n-n'+n''}H_B^*(2\omega_{n'}-\omega_{n''}-\omega_n)e^{2i(\omega_{n'}-\omega_{n''})t}.\label{NBb}
\end{align}
By using $H_B(\omega)=H_B^*(-\omega)$ and a change of indices, it is straightforward to show that $\mathcal{B}_n^{b}=-\mathcal{B}^f_{n}$. This implies that the modal equations (\ref{modal1}, \ref{modal1b}), derived from the propagation equations for the forward or backward fields, are identical, as one would expect. $\blacksquare$

 We insert the nonlinear terms Eqs. (\ref{KFn},\ref{NBf}) in Eq.
(\ref{modal1}) [equivalently Eqs. (\ref{KFn},\ref{NBb}) in Eq.
(\ref{modal1b}) ]
%\begin{align*}
% \beta_1 \dot a_n & -\left(\frac{\ln\rho_1\rho_2}{2L} - i\frac{\delta}{2L}\right)a_n+i\frac{\beta_2}{2}\left(\ddot a_n -2i\omega_n\dot a_n -\omega_n^2a_n\right)-\frac{\theta_1}{2L}E_{in}\delta_{n0}=
%\\&i\gamma \sum_{n',n''}\left(a_{n'}a_{n''}^*a_{n-n'+n''}  +Xa_{n'}a_{n''}^*a_{n+n'-n''}e^{-2i(\omega_{n'}-\omega_{n''})t}\right)+\\
%&i\frac{g_B}{2A_{eff}}\sum_{n',n''}a_{n'}a_{n''}a^*_{n-n'+n''}H_B(2\omega_{n'}-\omega_{n''}-\omega_n)e^{-i(2\omega_{n'}-\omega_n)t}.
%\end{align*}
%
%%
and assume that the modal amplitudes changes slowly over a round-trip, i.e. $|\dot a_n|\ll|\omega_n a_n|$. This assumption permits to simplify the dispersive contribution and to average out the fast oscillations in the  nonlinear terms as follows:
%
%\begin{widetext}
\begin{align}
\nonumber \dot a_n &+\left(\kappa + i\frac{\delta}{2\beta_1L}-i\frac{\beta_2}{2\beta_1}\omega_n^2\right)a_n-\frac{\theta_1}{2\beta_1L}E_{in}\delta_{n0}= \\
\nonumber & i\frac{\gamma}{\beta_1}\left(\sum_{n',n''}a_{n'}a_{n''}^*a_{n-n'+n''}+Xa_n\sum_{n'}|a_{n'}|^2\right)\\
\label{modal_fin}  &+i\frac{g_B}{2A_{eff}\beta_1}a_n\sum_{n'}|a_{n'}|^2H_B(\omega_{n}-\omega_{n'}),
\end{align}
%\end{widetext}
%
where we have defined the decay rate $\kappa=-\frac{\ln\rho_1\rho_2}{2\beta_1L}$.

%The modal equations (\ref{modal_fin}), which describe the temporal evolution of the slowly varying amplitudes of the cavity modes, represent a significant result of this work. 
%Unlike existing coupled-mode models, it consists of a single set of equations, with no need for additional variables describing the acoustic wave. Moreover, it fully captures all possible interactions among the modes via SBS. For example, SBS cascades are naturally included within Eqs. (\ref{modal_fin}), which is not the case in other existing models to our knowledge. These models often restrict investigations to a single mode driving SBS \cite{Bai2021,Zhang2024,Zhang2023,Nie2024}. Such simplifications significantly limit the applicability of these models to realistic configurations.

The modal equations~(\ref{modal_fin}), which govern the temporal evolution of the slowly varying amplitudes of the cavity modes, constitute a central contribution of this work. In contrast to existing coupled-mode models, our formulation relies on a single unified set of equations and does not require auxiliary variables to represent the acoustic wave. Importantly, it captures the full range of mode interactions mediated by SBS, including complex dynamics such as SBS cascades. These processes are inherently accounted for in Eqs.~(\ref{modal_fin}), unlike in previous models \cite{Bai2021,Zhang2024,Zhang2023,Nie2024}, which typically consider SBS driven by a single optical mode. Such assumptions considerably restrict the scope of those models when applied to realistic configurations.

%Restricting the analysis to scenarios where only a single mode drives SBS, as is often assumed, limits the model’s applicability. 
%ADD SOME COMMENTS HERE ON THE CMT FOR BRILLOUIN AND EXISTING MODELS. THE MEAN FIELD WHICH WILL BE DERIVED BELOW IS EQUIVALENT.
%THIS CAN ALSO BE STATED IN THE INTRO.

\subsection{Mean-field equation}
A fully equivalent model can be obtained form Eqs. (\ref{modal_fin}) by defining the slowly varying envelopes of the forward and backward fields, expressed as
\begin{align}
\psi_F(z,t)&=\sum_m a_m(t) e^{i\beta_1\omega_m z}=\sum_m a_m(t) e^{i \frac{\pi m}{L} z},\\
\label{FSB} \psi_B(z,t)&=\sum_m a_m(t)  e^{-i\beta_1\omega_m z}=\sum_m a_m(t) e^{-i \frac{\pi m}{L} z}.
\end{align}
The modal amplitudes can be calculated from the envelopes as follows
\begin{align}
\label{FcoefF} a_m(t)&=\frac{1}{2L}\int_{-L}^{L}\psi_F(z,t)e^{-ik_m z}dz\\
\label{FcoefB} &=\frac{1}{2L}\int_{-L}^{L}\psi_B(z,t)e^{ik_m z}dz.
\end{align}
The fields $\psi_{F,B}$ are defined on an extended interval $z\in[-L,L]$, are periodic in space of period $2L$ and satisfy $\psi_F(z,t)=\psi_B(-z,t)$. Thanks to this relation we can relate the fields in the 'nonphysical' cavity $-L<z<0$ to the real cavity $0<z<L$ to their counter-propagating counterparts \cite{Cole2018}. We develop the approach for the backward field envelope $\psi=\psi_B$, but an equivalent equation is obtained from the forward field too.
We multiply Eq. (\ref{modal_fin}) by $e^{-i\beta_1\omega_n z}$ and sum over $n$. By using
\begin{align*}
\frac{\partial \psi}{\partial t}&=\sum_m \dot a_m(t) e^{-i\beta_1\omega_m z},\\
\frac{\partial^n \psi}{\partial z^n}&=\sum_m (-i\beta_1 \omega_m)^n a_m(t)e^{-i\beta_1\omega_m z},
\end{align*}
we get
%\begin{align}
%\nonumber \frac{\partial \psi}{\partial t}+\left(\kappa+i\frac{\delta}{2\beta_1L}\right)\psi+i\frac{\beta_2}{2\beta_1^3}\frac{\partial^2 \psi}{\partial z^2}-\frac{\theta_1}{2\beta_1L}E_{in}=\\
%\nonumber i\frac{\gamma}{\beta_1}\left(|\psi|^2+\frac{X}{2L}\int_{-L}^L|\psi|^2 dz\right)\psi\\
%+i\frac{g_B}{2A_{eff}\beta_1}\sum_{n,n'}a_n|a_{n'}|^2H_B(\omega_{n}-\omega_{n'})e^{-ik_nz}.\label{LLElab}
%\end{align}
%%
%
%Equation (\ref{LLElab}) can be conveniently written as
%\begin{align}\label{LLE}
%\nonumber \frac{\partial \psi}{\partial t}+\left(\kappa+i\frac{\delta}{2\beta_1L}\right)\psi+i\frac{\beta_2}{2\beta_1^3}\frac{\partial^2 \psi}{\partial z^2}-\frac{\theta_1}{2\beta_1L}E_{in}=i\frac{\gamma}{\beta_1}\left(|\psi|^2+X\langle|\psi|^2\rangle\right)\psi\\
%+i\frac{\gamma_B}{\beta_1}\langle\psi\rangle\left[\langle\psi\varphi^*\rangle
%+\langle\psi^*\rangle\left(\varphi-\langle\varphi\rangle\right)\right],
%\end{align}
%
%\begin{widetext}
\begin{align}\label{LLEtime}
\nonumber T_r\frac{\partial \psi}{\partial t}=&-\left(\alpha+i\delta\right)\psi-i2L\frac{\beta_2}{2\beta_1^2}\frac{\partial^2 \psi}{\partial z^2}+\theta_1 E_{in}\\
\nonumber &+2i\gamma L\left(|\psi|^2+X\langle|\psi|^2\rangle\right)\psi\\
&+i\frac{g_B}{2A_{eff}}\psi*\left[h_B\cdot\left(\psi*\psi^*(-z)\right)\right],
\end{align}
%\end{widetext}
%
where $\langle.\rangle=\frac{1}{2L}\int_{-L}^L(.)dz$
denotes spatial average, $T_r=2\beta_1 L$ is the roundtrip time and $\alpha=\kappa T_r\approx 1-\rho_1\rho_2$ the overall losses.
We have defined the (periodic) convolution as
\begin{equation}\label{conv}
[\psi*h_B](z)=\int_{-L}^{L}\psi(\xi)h_B(z-\xi)d\xi
\end{equation}
and the  Brillouin response in the spatial domain as
\begin{align}\label{h_B}
h_B(z)=\sum_n\frac{H_B(k_n/\beta_1)}{2L}e^{-ik_nz}.
\end{align}
By using the properties of Fourier series, we see that $h_B(z)=\sum_nh(z-n2L)$ is the periodic replication of the inverse Fourier transform (in $z$) of the Brillouin response Eq. (\ref{Hb}):
\begin{equation}
H_B(k/\beta_1)=\int_{-\infty}^{\infty}h(z)e^{ikz}dz
\end{equation}
with
\begin{align}
\nonumber h(z)=&\frac{\beta_1\Omega_B\Gamma_B}{\sqrt{\Omega_B^2-(\Gamma_B/2)^2}}e^{-\frac{\Gamma_B}{2}\beta_1 z}\\
&\times\sin\left(\sqrt{\Omega_B^2-(\Gamma_B/2)^2}\,\beta_1z\right)u(z),
\end{align}
where $u(z)$ is the Heaviside step function.

Equation (\ref{LLEtime}) is the main result of this work. It represents a generalization of the FPLLE \cite{Cole2018} thanks to the inclusion of a Brillouin term. It can be efficiently solved with standard Fourier split-step methods. Indeed, the periodic convolutions Eq. (\ref{conv}) are efficiently calculated through fast Fourier transform algorithms. Moreover, it is more suited for analytical investigations than Eqs. (\ref{propF}-\ref{BC2}), as illustrated in the following section. The transverse coordinate in Eq. (\ref{LLEtime}) is space, as conventional in microresonator community. It can be mapped into a fast-time coordinate as customary in fiber optic community as described in \cite{Ziani2024}.

\section{Linear stability analysis}
Equation (\ref{LLEtime}) enables a straightforward analysis of the stability of homogeneous solutions against finite-wavelength perturbations, a task that is significantly more challenging in the CWE model due to its complexity. The CW solutions of Eq. (\ref{LLEtime}) satisfy:
\begin{equation}
\label{steady}\theta_1^2 P_{in}=P_s\left[\alpha^2+(\delta-2\gamma P_sL(1+X_{eff}))^2\right]
\end{equation}
where $P_s=|\psi_s|^2$ is the intracavity power, $P_{in}=|E_{in}|^2$ and $X_{eff}$  is the effective XPM coefficient defined before [see Eqs. (\ref{stazF},\ref{stazB})].
%The effective XPM coefficient is $X_{eff}=X+\gamma_B/\gamma$, with $$\gamma_B=\frac{g_B \Gamma_B}{2A_{eff}\Omega_B}$$
 We can assume, without loss of generality, $\psi_s$ to be real, which implies that the input field must be complex and satisfies the following equation:
\begin{equation}
\theta_1 E_{in}=\psi_s[\alpha+i(\delta-2\gamma P_s L (1+X_{eff}))].
\end{equation}
We consider a perturbed homogeneous solutions $\psi(z,t)=\psi_s+\eta(z,t)$. Assuming  $\eta\ll\psi_s$  small, we obtain 
\begin{align}
\nonumber T_r\frac{\partial \eta}{\partial t}=&-(\alpha+i\delta)\eta-i\frac{d_2}{2}T_r\frac{\partial^2 \eta}{\partial z^2}\\
\nonumber &+2\gamma P_sL\left((2+X)\eta +\eta^* + X<\eta+\eta^*> \right)\\
&+2\gamma_B P_sL\left(\frac{\Omega_B}{\Gamma_B}\eta*h_B +<\eta+\eta^*>\right),
\end{align}
where we have defined $d_2=\beta_2/\beta_1^3$. We now expand the perturbation over the cavity modes with time-varying amplitudes:
\begin{equation}
\eta(z,t)=\varepsilon_{n}^+(t)e^{ik_nz}+(       \varepsilon_{n}^-)^*(t)e^{-ik_nz}.
\end{equation}
For $n\neq 0$, the amplitudes of the modal perturbations obey
\begin{align}
\nonumber T_r\dot\varepsilon_{n}^+=&-(\alpha+i\delta)\varepsilon_{n}^++i\frac{d_2}{2}T_rk^2_n\varepsilon_{n}^+\\
\nonumber &+2i\gamma P_s L[(2+X)\varepsilon_{n}^++\varepsilon_{n}^-+X(\varepsilon_{n}^++\varepsilon_{n}^-)\delta_{n0}]\\
\label{evoep}&+ 2i\gamma_B P_sL[(\varepsilon_{n}^++\varepsilon_{n}^-)\delta_{n0} +\varepsilon_{n}^+\frac{\Omega_B}{\Gamma_B}H^*_B(\omega_n)],\\
\nonumber T_r\dot\varepsilon_{n}^-=&-(\alpha-i\delta)\varepsilon_{n}^--i\frac{d_2}{2}T_rk^2_n\varepsilon_{n}^-\\
\nonumber &-2i\gamma P_s L[(2+X)\varepsilon_{n}^-+\varepsilon_{n}^+ +X(\varepsilon_{n}^++\varepsilon_{n}^-)\delta_{n0}]\\
\label{evoep*} &- 2i\gamma_B P_sL[(\varepsilon_{n}^++\varepsilon_{n}^-)\delta_{n0} +\varepsilon_{n}^-\frac{\Omega_B}{\Gamma_B}H^*_B(\omega_n)].
\end{align}
%
%The last terms in the equations (\ref{evoep},\ref{evoep*}), which appears only for the zero mode, stem from the integral term which do not average to zero as in the case $n\neq0$. This difference is not present for the ring resonator, for which $X=0$. 
The system (\ref{evoep}-\ref{evoep*}) can be written as
 $d/dt(\varepsilon_{n}^+,\varepsilon_{n}^-)^T=M_n(\varepsilon_{n}^+,\varepsilon_{n}^-)^T$, and the eigenvalues of the matrix $M_n$ determine the stability of the solution.
 The temporal growth rate of the perturbations for $n\neq0$ reads :
 \begin{equation}\label{sigma}
\sigma(k_n)=\frac{1}{T_r}{\rm Re}\left[-\alpha+\sqrt{(2\gamma P_s L)^2-\mu_n^2}\right],
 \end{equation}
 where $$\mu_n=-\delta+2L\frac{\beta_{2}}{2}\,\omega_n^2+2 P_s L[(2+X)\gamma+\frac{g_B}{2 A_{eff}}H_B^*(\omega_n)].$$ 
 %$$H_B^*=\frac{\Omega_B^2}{\Omega_B^2-\omega_n^2+i\omega_n/\tau_B}.$$

 The temporal growth can be expressed as a spatial gain as $g(\omega_n)=\beta_1\sigma(k_n)$. If we  take ($\alpha=\delta=\gamma=d_2=0$) in Eq. (\ref{sigma}) we obtain the Brillouin gain function:
 \begin{align}
\nonumber  g_{Br}(\omega)=&-\frac{g_B}{2 A_{eff}}P_s{\rm Im}[H_B(\omega)]=
     \\\label{gain_bril} &-\frac{g_B}{2 A_{eff}}P_s\frac{\omega\Omega_B\Gamma_B^2}{(\Omega_B^2-\omega^2)^2+(\Gamma_B\omega)^2}.
 \end{align}

 Equation (\ref{gain_bril}) represents a double Lorentzian, with maximum gain (absorption) at negative (postitive) frequency shift $\mp \Omega_B$. It is consistent with Eqs. (\ref{PF}-\ref{gain_brillouin_2}), assuming CW operation and undepleted pump. 
  
As happens FP resonators featuring Kerr nonlinearity only, the growth rate of the $n=0$ mode cannot be derived from Eq. (\ref{sigma}) \cite{Cole2018,Ziani2024}. From the eigenvalues of the matrix $M_0$ we have:
 \begin{align}\label{sigma0}
&\sigma(0)=
\frac{1}{T_r}{\rm Re}\bigg[-\alpha+\\
\nonumber  &\sqrt{(2\gamma P_s L(1+X_{eff}))^2-(\delta-4(1+X_{eff})\gamma P_s L))^2}\bigg],
 \end{align}
 
It is straightforward to verify  that the condition $\sigma_0>0$ corresponds to the negative-slope branch of the multivalued response  $P_s(P_{in})$ derived from Eq. (\ref{steady}). This is consistent with the well-known scenario of optical bistability, in which branches with negative slope are typically unstable.

\section{Numerical results}
\begin{table}[b]
    \begin{ruledtabular}
    \begin{tabular}{ccc}
%    \hline
        \textbf{Parameter} & \textbf{Symbol} &\textbf{Value} \\
        \hline
        Cavity length & $L$ & 8.7472 cm\\
        %\hline
         Free spectral range & FSR &  1.176 GHz\\ 
         %\hline
        Cavity Finesse & $\mathcal{F}=\pi/\alpha$ & 420\\
       %  \hline
         Mirror power reflectivity &  $\rho^2$ & 0.998429\\
         %\hline
         Inverse group velocity & $\beta_1$ & 4.86 ns/m\\
         %\hline
          Group velocity dispersion & $\beta_2$ &  0.382 ps$^2$/km\\
          %\hline
      %   Third-order dispersion, $\beta_3$ & -0.00273  ps$^3$/km \\
         %\hline
         Kerr nonlinearity & $\gamma$ & 10.8 W$^{-1}$km$^{-1}$ \\
         %  \hline
        Effective area & $A_{eff}$ & 12.4 $\mu$m$^{2}$ \\
        %\hline
        Brillouin frequency shift & $\Omega_B=2\pi\nu_B$ & $2\pi\times$9.655 GHz \\
        %\hline
       Brillouin linewidth & $\Delta\nu_B=\frac{\Gamma_B}{2\pi}$ &  55 MHz \\
        % \hline
          Brillouin gain&  $g_B$ &  4.71$\times 10^{-12}$ m/W \\
        %\hline
        & $g_B/A_{eff}$ & 0.38 W$^{-1}$m$^{-1}$ \\
        %\hline
    \end{tabular}
    \end{ruledtabular}
    \caption{Values of the parameters used in the numerical simulations. Mirror transimissivity is obtained from $\theta_{1,2}^2=1-\rho^2$ considering ideal mirrors. The effective values of $\rho_{1,2}$ are instead calculated from finesse \citep{Ziani2024}, which includes all possible sources of losses. }
    \label{parameters}
\end{table}
Equation (\ref{LLEtime}) is solved using the split-step Fourier scheme in conjunction with the fourth-order Runge-Kutta method for the nonlinear terms. The periodic convolutions in the last term of Eq. (\ref{LLEtime}) are efficiently computed using the fast Fourier transform (FFT) algorithm. In contrast, Eqs. (\ref{propF}-\ref{BC2}) are solved with an operator splitting method: the equation involving both propagation and dispersion is evaluated using a finite-difference predictor–corrector scheme with integration along the characteristics ($\Delta z/ \Delta t=v_g$, with $\Delta t$ and $\Delta z$ the time and space grid steps), while the nonlinear terms are addressed with the fourth-order Runge-Kutta method \cite{Sun2019, Dong2016}. It is important to note that the counter-propagation described by the advective terms imposes a relationship between the time and space increments to ensure a stable numerical scheme (Courant-Friedrichs-Lewy condition \cite{LeVeque_2002}).
To demonstrate the accuracy and efficiency of the numerical solution for the mean-field model Eq. (\ref{LLEtime}), we present examples of comb generation in a fiber FP resonator. The parameters, corresponding to a high-finesse FP cavity made of a highly nonlinear fiber \cite{Bunel2024SW,postdeadline,BunelNatPhot}, are summarized in Table \ref{parameters}.

\begin{figure}
\includegraphics[width=0.49\columnwidth]{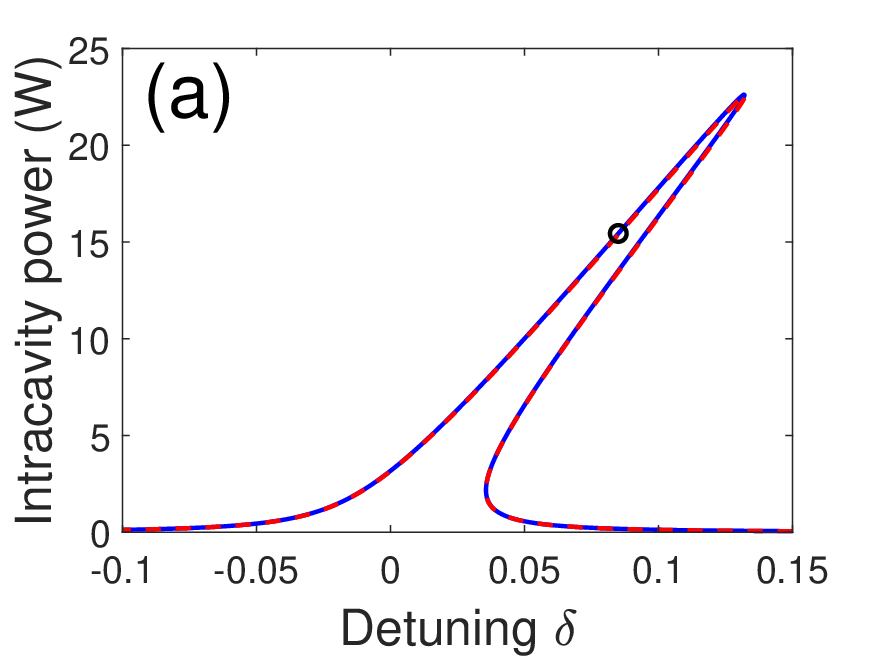}
\includegraphics[width=0.49\columnwidth]{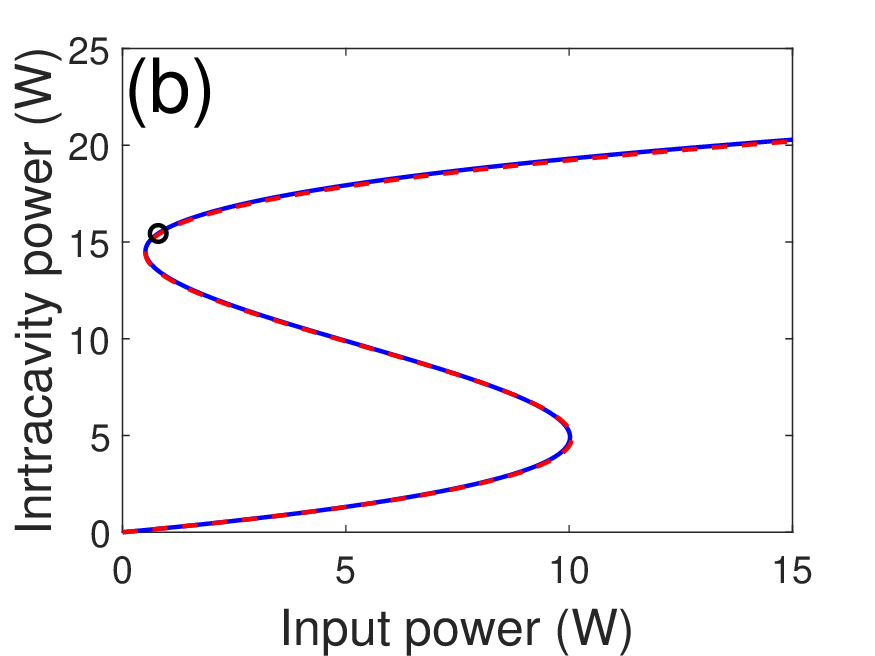}   
\includegraphics[width=0.49\columnwidth]{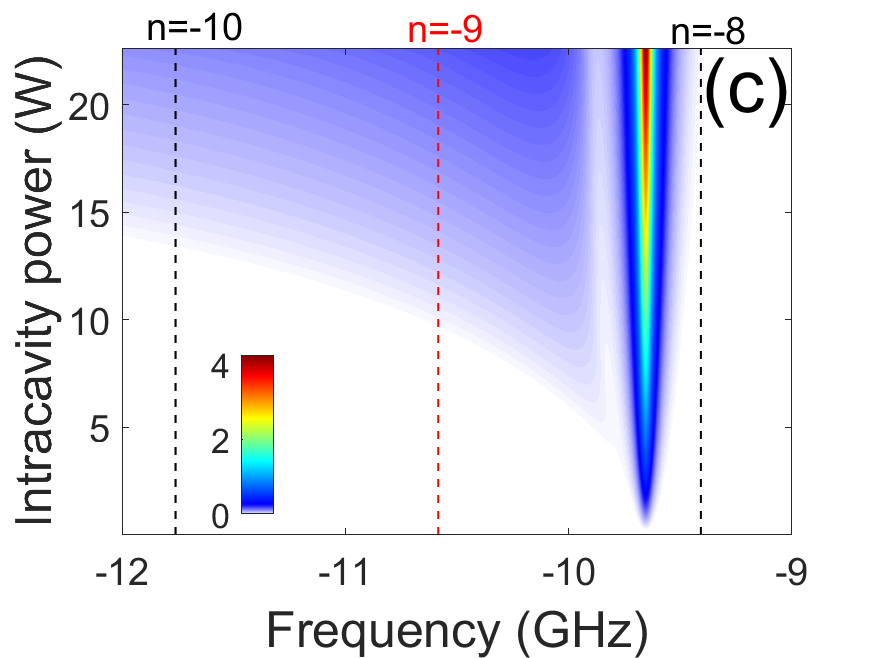}  
\includegraphics[width=0.49\columnwidth]{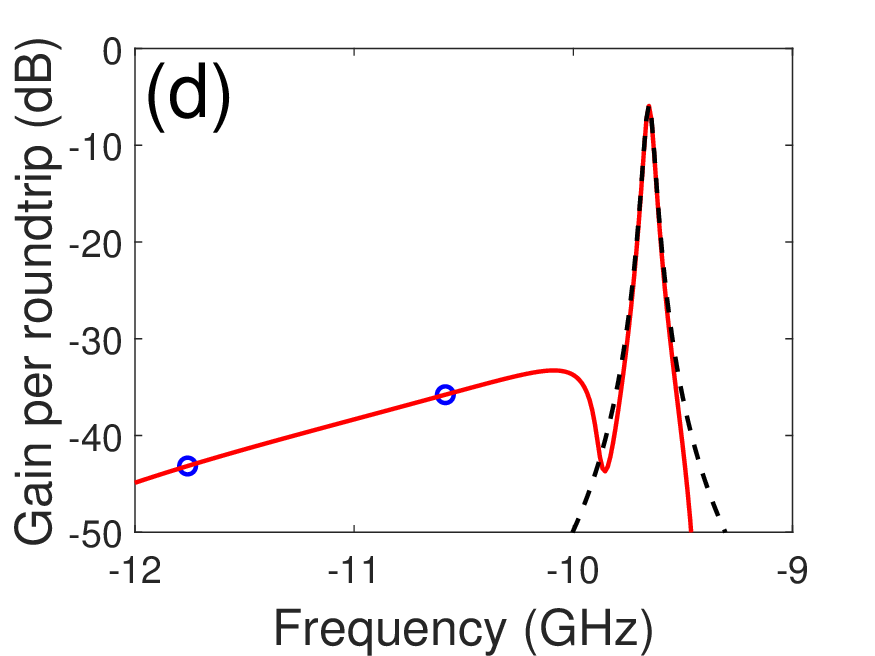}  
\caption{Steady state curves for CWE Eq. (\ref{eq:SS_power}) (solid blue curve) and mean field model Eq. (\ref{steady}) (dashed red curve) as a function of (a) detuning for fixed $P_{in}=0.8$ W and (b) input power for a fixed detuning $\delta=0.085$. (c) False-color plot of the parametric gain for perturbations on the upper branch of the steady state response for $P_{in}$=0.8 W. Dashed lines indicate the cavity modes. (d) Gain profile taken at $(P_{in},\delta)=(0.8{\rm W},0.085 )$, corresponding to the
 black circles in (a,b) (the corresponding intracavity power is $P=15.44$ W). Blue dots specify cavity modes $n=8,9$ which experience gain. Black dashed curve, pure Brillouin gain from Eq. (\ref{gain_bril}). \label{fig_SS}}
\end{figure}

Figure \ref{fig_SS} shows the steady state of the cavity calculated via CWE (solid blue curve) and mean field model (dashed red curve) for fixed input power $P_{in}=0.8$ W (a), and for fixed detuning $\delta=0.085$ in the bistable regime (b).
The two curves are nearly superimposed, demonstrating that the hypotheses for the validity of the mean field limit are satisfied. Additionally, the introduced Brillouin response accurately models the effect of SBS on the continuous-wave pump. It has been shown that wide frequency combs can be generated through Kerr and Brillouin effect in normal dispersion resonators thanks to a particular parametric process even when the peak of the Brillouin gain does not overlap with a cavity resonance \cite{postdeadline,BunelNatPhot}. Figure \ref{fig_SS}(c) shows the parametric gain $g(\omega)$ from Eq. (\ref{sigma}) as a function of frequency (the mode number is considered  as continuous variable) and intracavity power (equivalently detuning) on the upper branch of the steady response. The parametric gain is symmetric in $\omega$, so only the negative frequencies are shown. The strongest peak corresponds to the usual Brillouin gain, which in this configuration  falls in between two cavity modes ($n=-8$ and $n=-9$). The system is thus stable if only Brillouin gain is considered. However, the interplay with Kerr nonlinearity induces an additional, broader lobe, which can excite several modes, the most unstable being $n=-9$ in our example (see dashed red line in Fig. \ref{fig_SS}(c)). A slice of the parametric gain corresponding to $(P_{in},\delta)=(0.8{\rm W},0.085 )$ is reported in Fig. \ref{fig_SS}(d), red curve. Dashed-black curve shows the pure Brillouin gain from Eq. (\ref{gain_bril}), and the cavity modes $n=-8,-9$ are indicated by the blue dots. 
Perturbations of the CW state with a frequency corresponding to $\pm9$ FSR grow from noise and can give rise to frequency combs through a four-wave mixing cascade. The generation dynamics is described in the following examples.

%Figure \ref{smallcomb} reports the simulated generation of a quite narrow OFC. Panel (a) shows the evolution of the backward field power $|B|^2$ as a function of rountrips calculated from CWE. We see the generation of a stable temporal pattern drifting towards the left. The simulation of mean field model gives a nearly identical resuts (not shown). A comparison of the two models is reported in panel (c), where the intracavity power at rountrip 8000 is shown. We report the backward field $B$ (red curve) and the reversed forward field $F(-z)$ (blue curve), which can directly compared with the mean field $\psi$ (dashed black curve). We calculate the spectrum following the definition Eq. (\ref{FSB}) and relate the mode numbers to frequencies thanks to the dispersion relations Eq. (\ref{modes}). The spectral evolution from CWE reported in panel (b) shows the birth of two symmetric lines at 9 FSR from the pump as predicted by linear stability analysis. This initial modulation generates a comb several GHz wide thank to cascaded FWM. Panel (d) reports the comparison of the spectra after 8000 roundtrips from CWE (red circles) and mean field (black crosses): the agreement is perfect, even for spectral components which are 150 dB less intense than the main CW line. In both models we used the same step $\Delta z$, which determines the size of the simulable spectral window. For 2048 modes for CWE was 140 min, whereas for mean field 3.7 s only, corresponding to a speed-up factor of 2000.

\begin{figure}
\includegraphics[width=0.49\columnwidth]{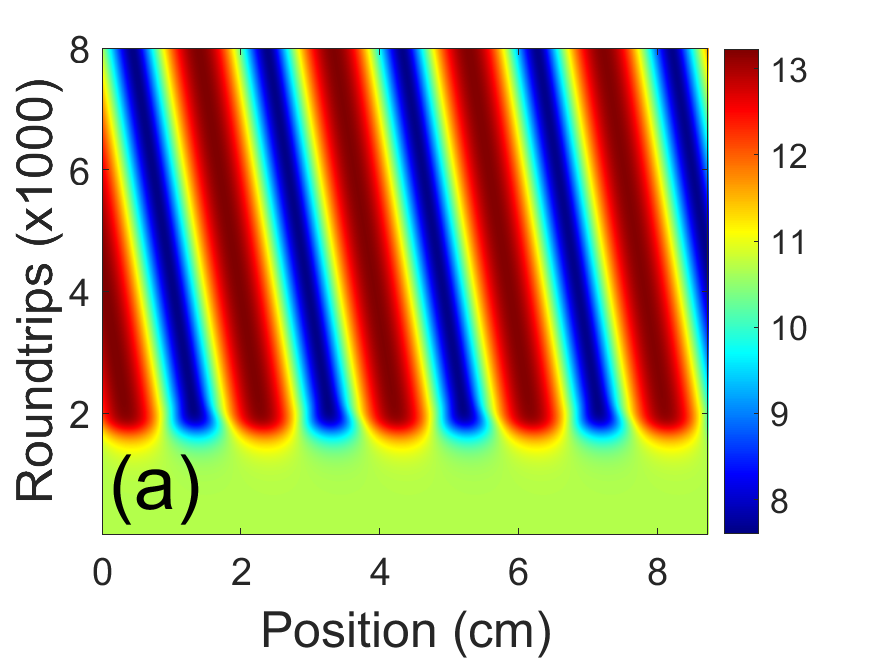}
\includegraphics[width=0.49\columnwidth]{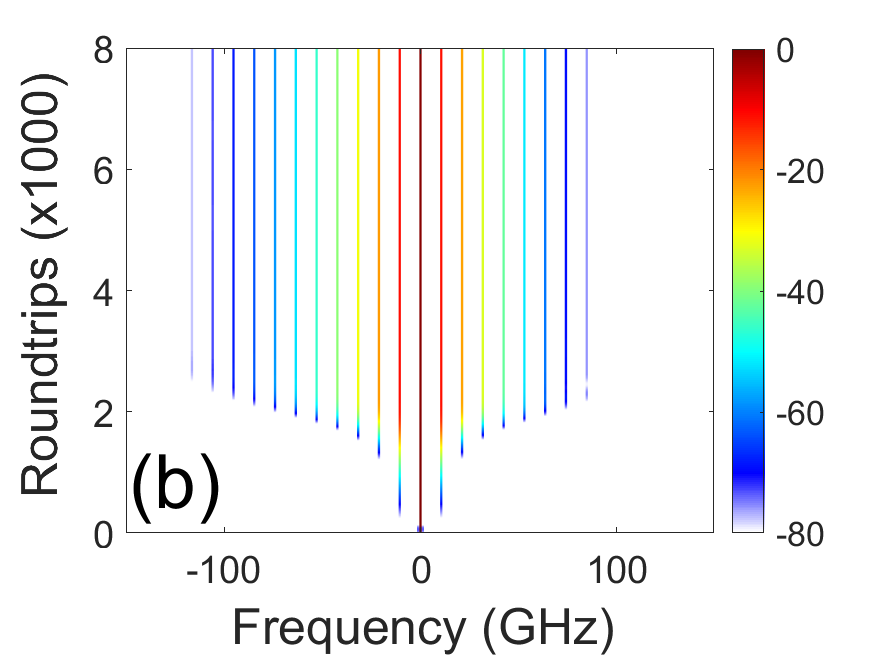}
\includegraphics[width=0.49\columnwidth]{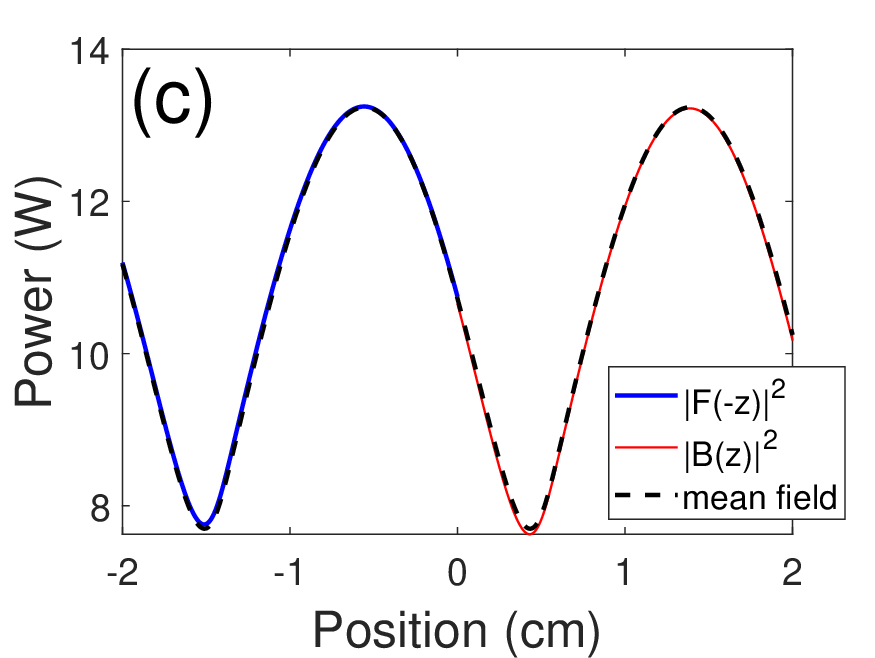}
\includegraphics[width=0.49\columnwidth]{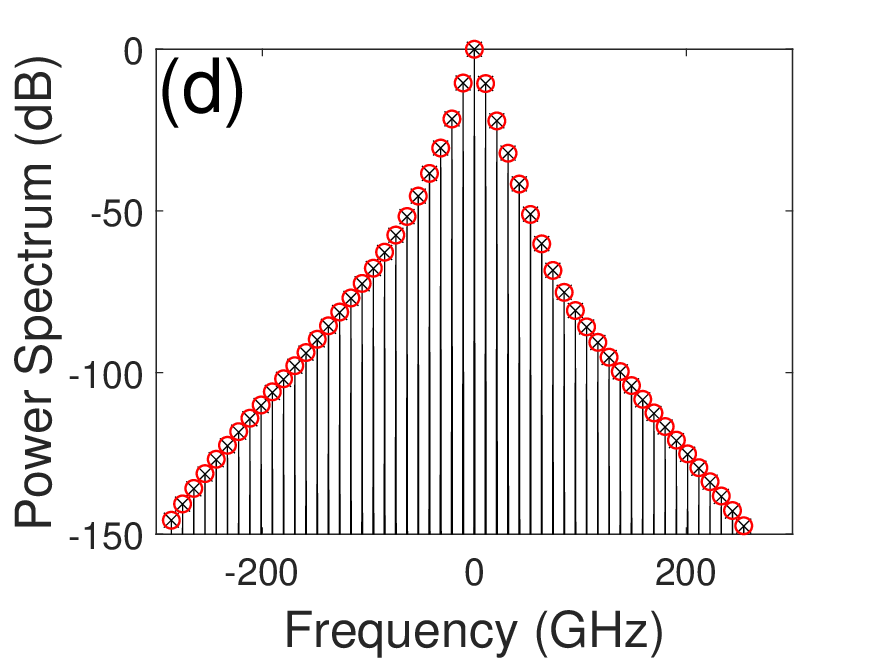}
\caption{(a) False color plot of intracavity power of backward field as a function of position and rountrips. (b) Evolution of the intracavity spectrum. (c) Comparison of field profiles at rountrip 8000 from CWE (blue curve, forward field; red curve, backward field) and mean-field (dashed black curve). (d) Comparison of spectra at rountrip 8000 from CWE (red circles) and mean field (black crosses). Computational time for 2048 spatial modes: CWE 140 min; mean field 3.7 s (2000 times faster).  Operating point: $P_{in}=0.8$ W, $\delta=0.055$ \label{smallcomb}}
\end{figure}

Figure \ref{smallcomb} illustrates the simulated generation of a relatively narrow OFC with $P_{in}=0.8$ W and $\delta=0.055$. Panel (a) shows the evolution of the backward field power $|B|^2$ as a function of roundtrips, calculated from CWE. The field is displayed at times corresponding to multiples of the roundtrip time. This sampling removes the rapidly oscillating temporal phase from the fields $F,B$, enabling clearer visualization and direct comparison with the mean-field variables $\psi_F$ and $\psi_B$ [cfr. Eqs. (\ref{expanb}) and (\ref{FSB})]. A stable temporal pattern emerges, drifting towards the left. Simulations based on the mean field model yield nearly identical results (not shown). 
Panel (c) provides a direct comparison between the two models, showing the intracavity power at roundtrip 8000, well after that a steady-state has been reached. The backward field $B$ (red curve) and the reversed forward field $F(-z)$ (blue curve) are compared with the mean field $\psi$ (dashed black curve), demonstrating excellent agreement.
The spectral evolution from CWE, reported in panel (b) is obtained by performing Fourier transform of $F(L,t)$ on a time window of duration $T_r$ at successive roundtrips. It reveals the emergence of two symmetric lines, 9 FSR from the pump, as predicted by linear stability analysis. This initial modulation grows into a comb spanning several GHz, driven by cascaded FWM. Panel (d) compares the spectra after 8000 roundtrips for CWE (red circles) and the mean field model (black crosses). For the mean field, the spectrum is computed using the definition in Eq. (\ref{FcoefB}), with mode numbers translated into frequencies via the dispersion relations in Eq. (\ref{modes}). 
The agreement is remarkable, extending even to spectral components 150 dB weaker than the primary CW line.
Both models used the same spatial step $\Delta z$, which determines the number of modes and the size of the spectral window that can be simulated. For 2048 modes, the CWE simulation required 140 minutes, while the mean field model completed in just 3.7 seconds, achieving a speed-up factor of 2000.

It is worth emphasizing that, unlike the majority of Kerr frequency comb generation in passive cavities where the repetition rate is fixed by the cavity FSR, the process described here allows comb generation with multiple-FSR repetition rates. The repetition rate is primarily governed by the parametric gain frequency, which, in this example, reaches its maximum at 9 times the FSR, rather than being directly determined by the cavity length.

\begin{figure}
\includegraphics[width=0.49\columnwidth]{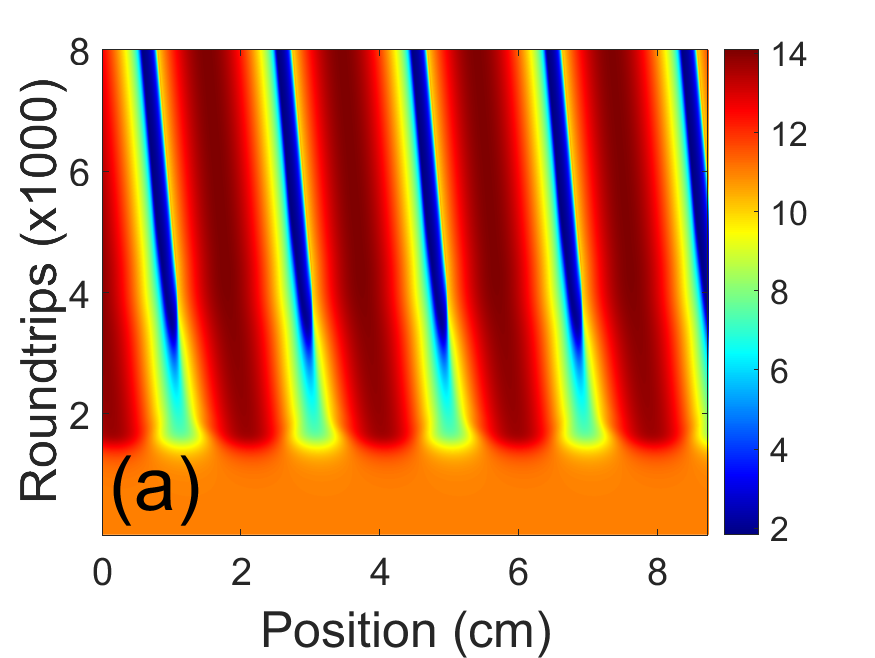}
\includegraphics[width=0.49\columnwidth]{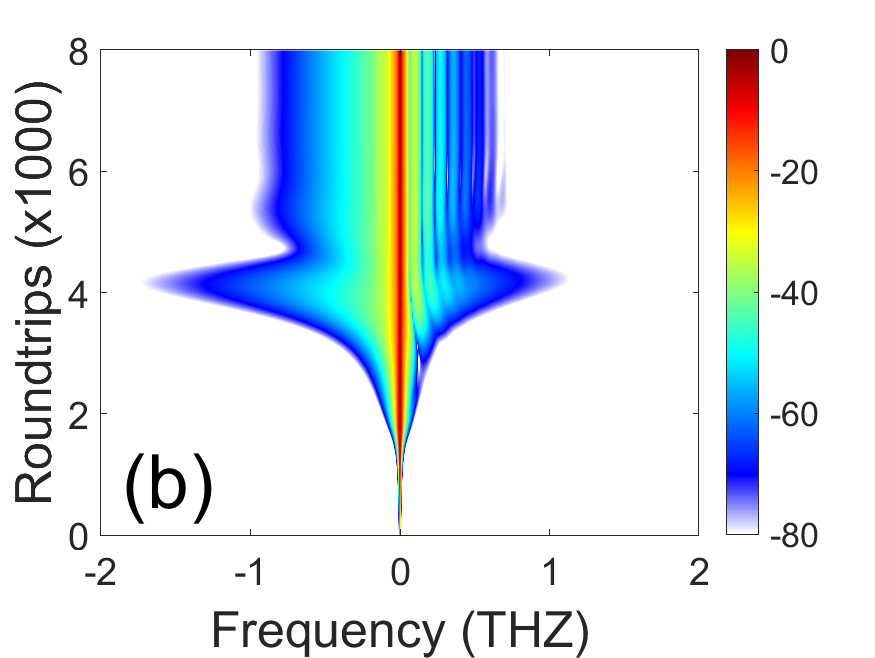}
\includegraphics[width=0.49\columnwidth]{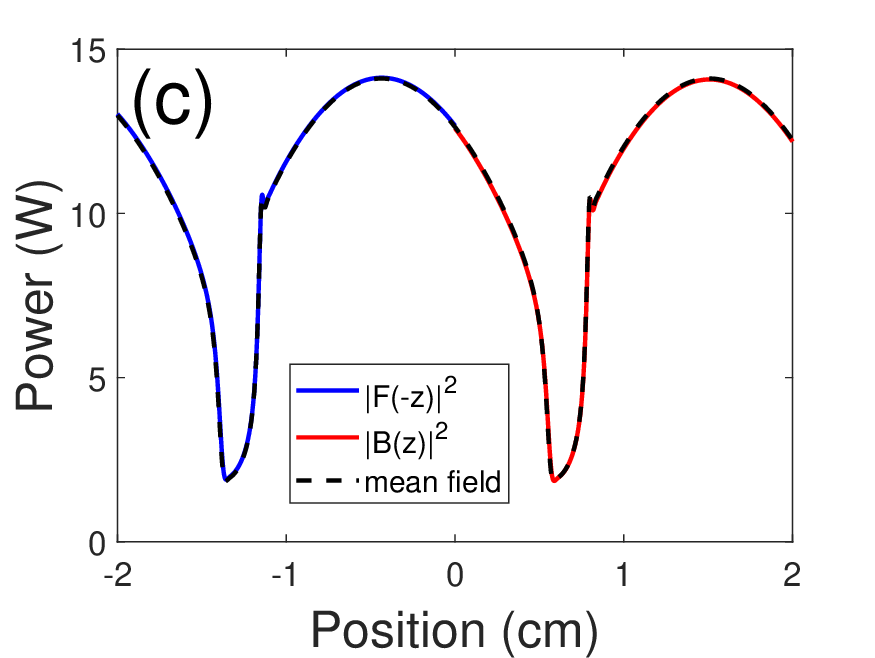}
\includegraphics[width=0.49\columnwidth]{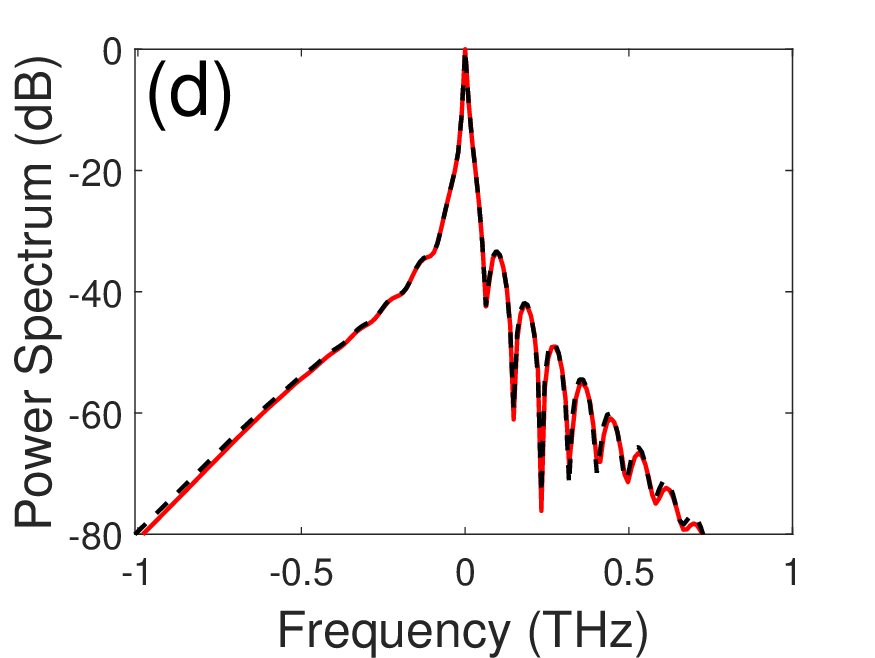}
\caption{As in Fig. \ref{smallcomb}. In panel (b) and (d) only the spectral components at multiples of $9 \times {\rm FSR}$ are considered in order to achieve a smooth representation (CWE: solid red, mean field: dashed black).   Computational time for 8092 spatial modes: CWE 1327 min; mean field 10.4 s (8000 times faster).  Operating point: $P_{in}=0.8$ W, $\delta=0.057$ \label{mediumcomb}}
\end{figure}

Figure \ref{mediumcomb} illustrates the generation of a broader frequency comb, exceeding 1.5 THz at -80 dB, achieved by slightly increasing the detuning to $\delta=0.057$. Panel (a) shows the generation of a pattern drifting to the left, while panel (c) reveals a highly asymmetric temporal waveform characterized by a steep leading front with a small spike and a smoother trailing front. This corresponds to the broad spectrum displayed in panels (b) and (d). For clarity, only spectral peaks spaced by nine times the free spectral range are shown, resulting in a smoother representation. The remaining cavity modes are minimally excited, with amplitudes approximately 100 dB lower than the primary components. The agreement between the CWE and mean-field models is remarkable, as evidenced by their nearly superimposed traces in both the temporal (c) and spectral (d) domains. To accurately capture the spectral extent of the comb, 8092 spatial modes were used. The corresponding computational times were 1327 minutes for the CWE model and 10.4 seconds for the mean-field model, demonstrating a speed improvement of approximately 8000 times.

\begin{figure}
\includegraphics[width=0.49\columnwidth]{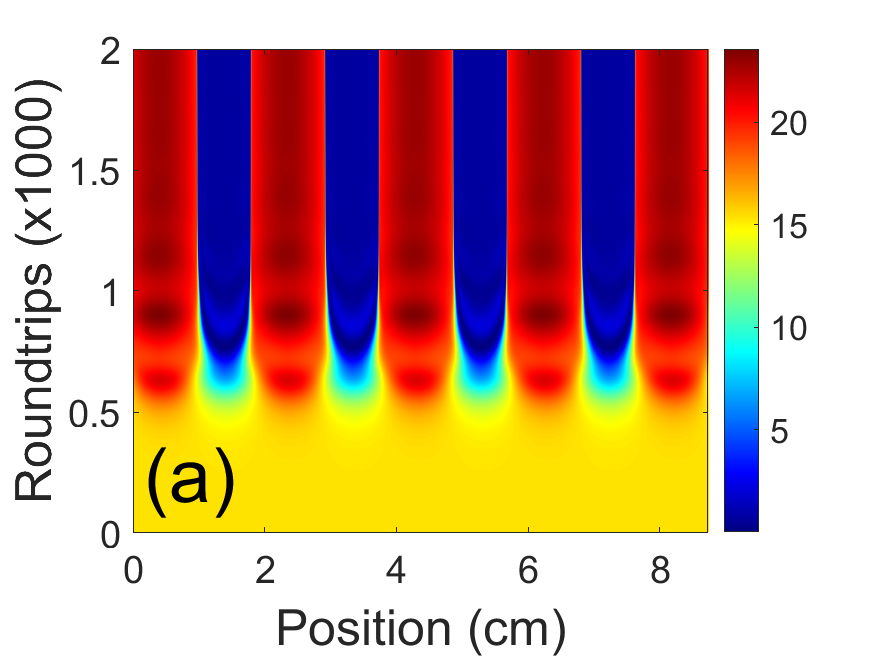}
\includegraphics[width=0.49\columnwidth]{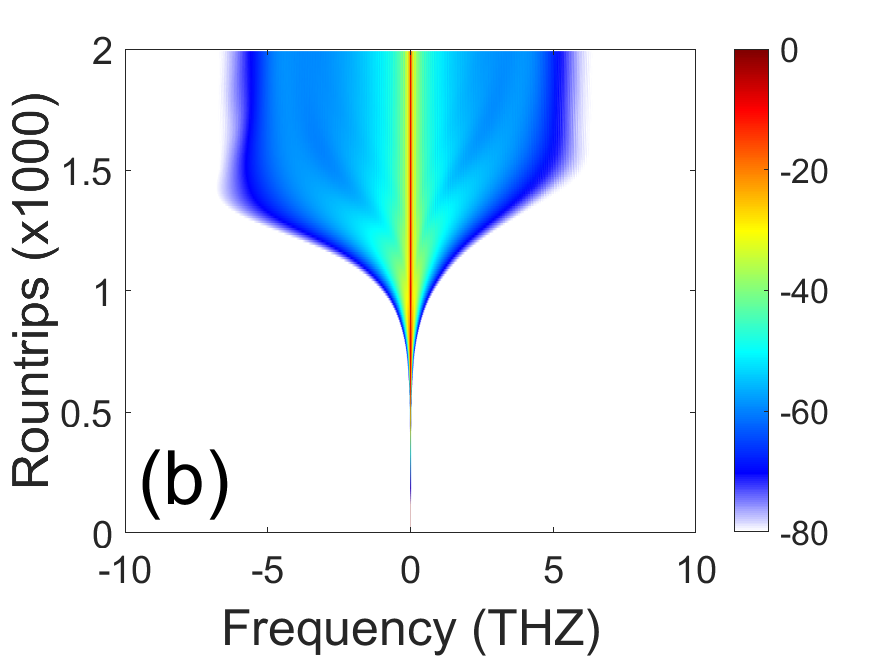}
\includegraphics[width=0.49\columnwidth]{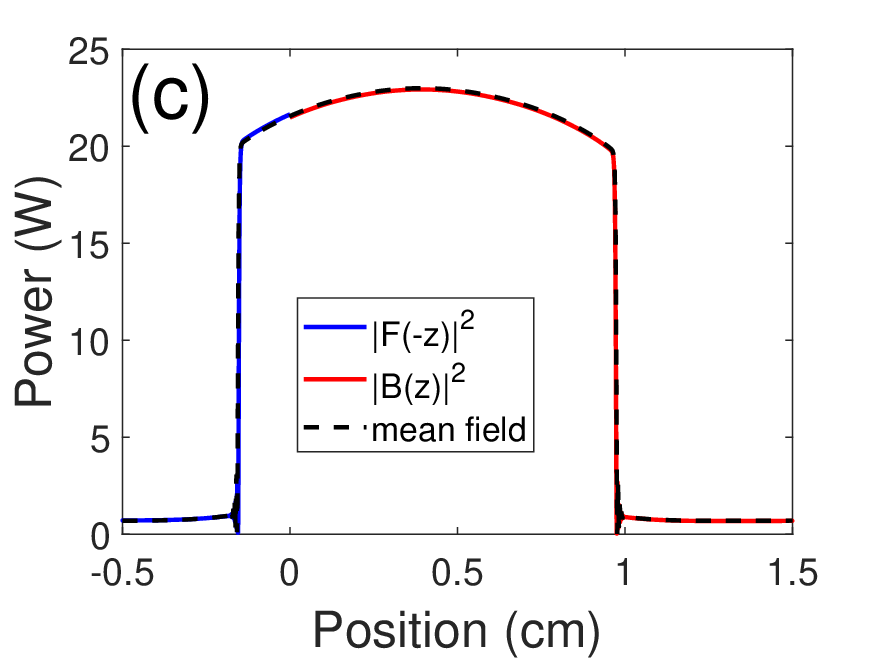}
\includegraphics[width=0.49\columnwidth]{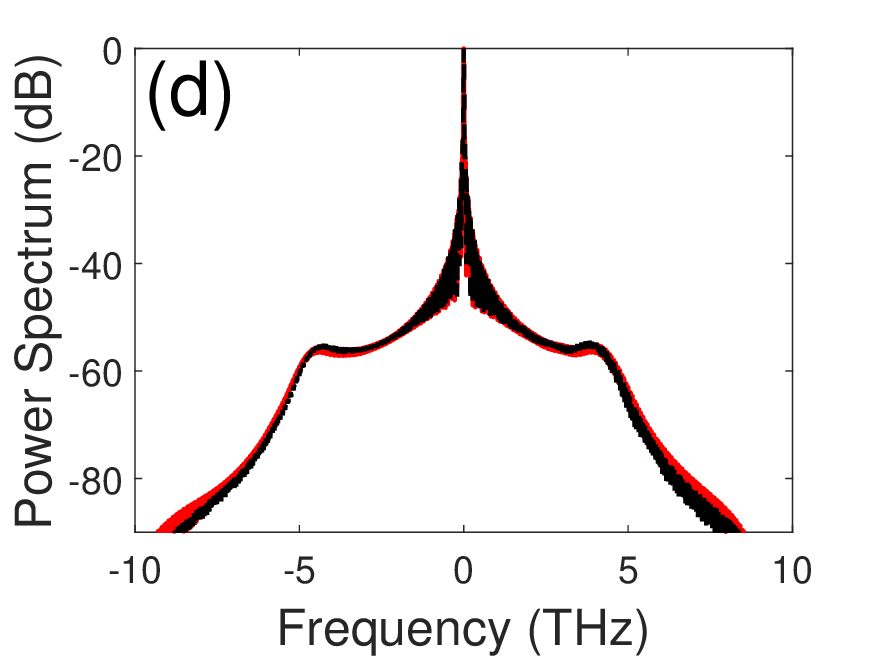}
\caption{As in Fig. \ref{mediumcomb}. Computational time for $2^{15}$ spatial modes: CWE 6716 min; mean field 42 s (9600 times faster).  Operating point: $P_{in}=0.8$ W, $\delta=0.085$ \label{largecomb}}
\end{figure}

By increasing the detuning to  $\delta=0.085$, an even larger comb is generated, exceeding 20 THz at -80 dB, as shown in Fig. \ref{largecomb}. The temporal evolution in panel (a) demonstrates the formation of a stable, non-drifting waveform. The individual pulses are  symmetric, characterized by steep leading and trailing fronts that transition between a high, nearly flat state and a lower level, as depicted in panel (c). The corresponding spectrum exhibits a sinc-like profile centered around the pump, with two prominent shoulders at $\pm$ 5 THz. These temporal and spectral characteristics indicate the formation of switching waves, a phenomenon known to sustain normal dispersion comb generation \cite{Macnaughtan2023, Bunel2024SW, Anderson2022}. The absence of drift suggests that the pairs of switching waves forming the nine bright pulses circulating in the resonator are excited at the Maxwell point \cite{Parra-Rivas2016}. Remarkably, even under these extreme conditions —featuring steep pulse fronts and a broad spectral range— the mean-field model accurately reproduces the results obtained from the CWE model, as evidenced by the nearly superimposed curves in panels (c) and (d). To fully capture the spectral extent of the comb, $2^{15}$ spatial modes were employed. The computational times were 6716 minutes for the CWE model and just 42 seconds for the mean-field model, representing a speed improvement of 9600, nearly four orders of magnitude.

\section{Conclusion}
We have introduced a novel mean-field equation to describe the nonlinear dynamics of a Fabry-Perot resonator containing a dispersive medium, incorporating both Brillouin and Kerr nonlinearities. %Our model generalizes the Lugiato-Lefever equation for Fabry-Perot resonators by including a properly defined Brillouin response. %Through this framework, we demonstrated the ability to describe the generation of Kerr-Brillouin combs in the normal dispersion regime.
Our results highlight the model's capability to accurately reproduce the outcomes of coupled wave equations for optical and acoustic fields with exceptional precision. Moreover, our approach achieves a significant reduction in computational time—up to four orders of magnitude—compared to simulations of the coupled wave equations. The simple yet robust mathematical structure of the proposed model also facilitates analytical insights, such as the evaluation of the stability and growth rates of homogeneous solutions.
An illustration has been provided here in the normal dispersion regime to highlight the precision of the model and the time savings it offers, but it obviously apply to anomalous dispersion as well.
We believe that this framework will serve as a powerful tool for the study and design of frequency combs arising from the interplay of Kerr and Brillouin nonlinearities, advancing both fundamental understanding and practical applications in this domain.
 This could open new avenues of investigation in comb generation by considering the Brillouin effect as an advantage or an additional degree of freedom, rather than attempting to cancel or avoid it (as seldom done in nonlinear fiber optics devices).

\begin{acknowledgments}
M.C. acknowledges fruitful discussions with F. Leo and S-P. Gorza.
%The present research was supported by the agence Nationale de la Recherche (Programme Investissements d’Avenir, ANR FARCO, ANR TRIPLE, LABEX CEMPI); Ministry of Higher Education and Research; European Regional Development Fund (WAVETECH), the CNRS (IRP LAFONI); Hauts de France Council (GPEG project), and the university of Lille (LAI HOLISTIC) % Create the reference section using BibTeX:
%Illuminating discussions with A. Mussot, T. Bunel and A.M. Perego are gratefully acknowledged.
\end{acknowledgments}

\providecommand{\noopsort}[1]{}\providecommand{\singleletter}[1]{#1}%


\begin{thebibliography}{57}
\bibitem [{env()}]{envelopes}%
  \BibitemOpen
  \href@noop {} {}\bibinfo {note} {This factor stems from a different
  definition of the complex envelopes, namely (i) $E=(Ae^{i\omega t} + c. c.)$
  vs (ii) $E=(A/2e^{i\omega t} + c. c.)$. It is important to note that
  convention (i) is followed in Boyd's textbook, while convention (ii) is used
  by Agrawal. This distinction affects how the nonlinear coefficients are
  defined.}\BibitemShut {Stop}%
\end{thebibliography}
\end{document}